\documentclass[preprint,authoryear,12pt]{elsarticle}
\usepackage{natbib}

\usepackage{graphicx}
\usepackage{amssymb}
\usepackage[ps2pdf,%
a4paper=true,%
breaklinks=true,%
colorlinks=true,%
pdfauthor={First Author et al.},%
pdftitle={Template for manuscripts in Advances in Space Research}%
]{hyperref}
\journal{Advances in Space Research}

\begin{document}

\begin{frontmatter}

\title{Discovery of New X-ray Sources near the Unidentified Gamma-ray Source HESS~J1841-055}

\author[kyoto]{K. K. Nobukawa\corref{cor}}
\ead{kumiko@cr.scphys.kyoto-u.ac.jp}
\author[hakubi,kyoto]{M. Nobukawa}
\author[kyoto]{T. G. Tsuru}
\author[osaka,kyoto]{K. Koyama}

\address[kyoto]{Department of Physics, Graduate School of Science, Kyoto University, Kitashirakawa Oiwake-cho, Sakyo-ku, Kyoto 606-8502, Japan}
\address[hakubi]{The Hakubi Center for Advanced Research, Kyoto University, Yoshida-Ushinomiya-cho, Sakyo-ku, Kyoto 606-8302, Japan}
\address[osaka]{Department of Earth and Space Science, Graduate School of Science, Osaka University, 1-1 Machikaneyama-cho, Toyonaka, Osaka 560-0043, Japan}

\begin{abstract}
HESS~J1841$-$055 is a diffuse unidentified gamma-ray source with the size of
$\sim\,1.3^{\circ}\times\,1^{\circ}$. No conclusive counterpart in other wavelengths has so far detected. 
To search for X-rays responsible for the TeV emission, the Suzaku observations were conducted, which covered  a half region of the HESS source. 
In the soft band (0.5--2.0~keV), we discovered a diffuse emission, Suzaku~J1840.2$-$0552, with the size of $\sim10'$.
Since its spectrum was fitted by an optically thin thermal plasma model, Suzaku~J1840.2$-$0552 is likely to be a supernova remnant.  
We also discovered an extended source, Suzaku~J1840.2$-$0544, in the hard band (2.0--8.0~keV) with an emission line at 6.1~keV. 
From the spectral feature and large interstellar absorption, this source is likely to be a cluster of galaxies behind the 
Galactic plane at the red-shift of $\sim$0.09.
The other diffuse source spatially overlaps with the SNR candidate G26.6$-$0.2, which shows a non-thermal dominant spectrum.
Since no other candidate is found in the hard X-ray band, we infer that these largely extended sources could be possible counterparts of HESS J1841$-$055. 

\end{abstract}


\end{frontmatter}

\parindent=0.5 cm
\section{Introduction}

HESS~J1841$-$055 is a largely extended TeV gamma-ray source with the size of $\sim\,1.3^{\circ}\times\,1^{\circ}$ discovered by H.E.S.S. in the Galactic plane survey \citep{aharonian08}. 
With the ARGO-YBJ experiment, \cite{bartoli13} also detected a TeV gamma-ray emission from the position which is coincident with HESS~J1841$-$055. 
{\it Fermi}-LAT detected GeV gamma-rays from the HESS~J1841-055 region \citep{neronov10, Ac13}.

\cite{aharonian08} suggested that {\bf two sources} are possible counterparts, at least partly, of HESS~J1841$-$055. One is G\,26.6$-$0.1, which was detected by {\it ASCA} \citep{bamba03}.    
\cite{bamba03} assumed that G\,26.6$-$0.1 is a non-thermal supernova remnant (SNR) due to the hard X-ray spectrum (the photon index of $\Gamma =1.3$).
The other is AX~J1841.0$-$0536, a supergiant fast X-ray transient (SFXT; \citealt{neg06, sguera05}). AX~J1841.0$-$0536 exhibits many flares with time-scales from a few hundred seconds to a few hours with the large flux range of 
$\sim10^{-13}$--$10^{-10}$~erg s$^{-1}$~cm$^{-2}$ \citep{Romano09, Bozzo11, nobukawa12}.   
\cite{sguera09} also suggested that AX~J1841.0$-$0536 could be 
responsible for at least a fraction of the entire gamma-ray emission.
However, no conclusive fact to support the origin for the full emission from HESS~J1841$-$055 in any other wavelengths has been found.

In order to search for counterparts of HESS~J1841$-$055 in X-ray, we investigated the region around HESS~J1841$-$055 with {\it Suzaku}.
Although {\it Suzaku} covered only the southern half region of HESS~J1841$-$055,
we found three diffuse sources. One is a soft and largely diffuse source with a thermal spectrum. Another is small but 
has an extended hard X-ray emission with the emission line at 6.1~keV. The other is a largely extended source with a non-thermal spectrum. 
In this paper, we mainly report these new sources and discuss the possible relation to the gamma-ray emission.

\section{Observation and Data reduction}
We performed observations in the HESS~J1841$-$055 region with the X-ray Imaging Spectrometer (XIS; \citealt{koyama07}) 
onboard the {\it Suzaku} satellite \citep{mitsuda07} in March 2011 (ObsIDs=505088010, 505089010, and 505090010). 
We also used the archive data of ObsID$=$504052010 (PI, N. Kawai).
The observation information is listed in table~\ref{table1}.

The XIS consists of four X-ray CCD cameras which cover the energy range of 0.2--12~keV. Three of the cameras (XIS~0, 2, and 3) have front-illuminated 
devices and the remaining one (XIS~1) contains a back-illuminated device.
All of the cameras are placed on the focal planes of the X-Ray Telescopes (XRTs; \citealt{serlemitsos07}).
The whole region of XIS~2 and one-fourth of XIS~0 have been out of function since 2006 November and 2009 June, respectively, 
and hence the data from the remaining devices were used. 
The XIS was working with the normal clocking mode without window option during all observations.

We used the software packages HEAsoft v6.15.1 and XSPEC v12.8 for the data reduction and analysis. 
The XIS event files were converted to pulse-invariant channels using the {\tt xispi} software and the calibration 
database (CALDB) version 2013-12-31. We excluded events during the South Atlantic Anomaly passages, at elevation angles below 
5$^\circ$ from the night Earth rim, and at elevation angles below 20$^\circ$ from the sunlit Earth rim. Response file 
({\tt arf}) and redistribution file ({\tt rmf}) for spectral fittings were produced 
by {\tt xissimarfgen} \citep{ishisaki07} and {\tt xisrmfgen}, respectively. The non-X-ray background (NXB) for the XIS 
was generated by {\tt xisnxbgen} \citep{tawa08} and was subtracted from the data before analyses in the following sections.

Errors in this paper are quoted at a 90\% confidence level unless otherwise specified.

\begin{table}
\begin{center}
\caption{Observation log.}
\begin{tabular}{llll}
\hline
Obs. ID & Obs. Date & (R.A., Dec.)$_{\rm J2000}$	& Exposure time\\
\hline
504052010	& 2009-Apr-13	& ($279^{\circ}.7750$, $-5^{\circ}.7085$) & 41.1 ks\\
505088010	& 2011-Mar-25	& ($279^{\circ}.8307$, $-5^{\circ}.8897$) & 49.7 ks\\
505089010	& 2011-Mar-26	& ($280^{\circ}.1473$, $-5^{\circ}.9564$) & 50.0 ks\\
505090010	& 2011-Mar-27	& ($280^{\circ}.1471$, $-5^{\circ}.6090$) & 49.6 ks\\
\hline
\end{tabular}
\label{table1}
\end{center}
\end{table}

\section{Result \& Discussion} 
\subsection{X-ray image} 

\begin{figure}[htpb]
\begin{center}
\includegraphics[width=14cm]{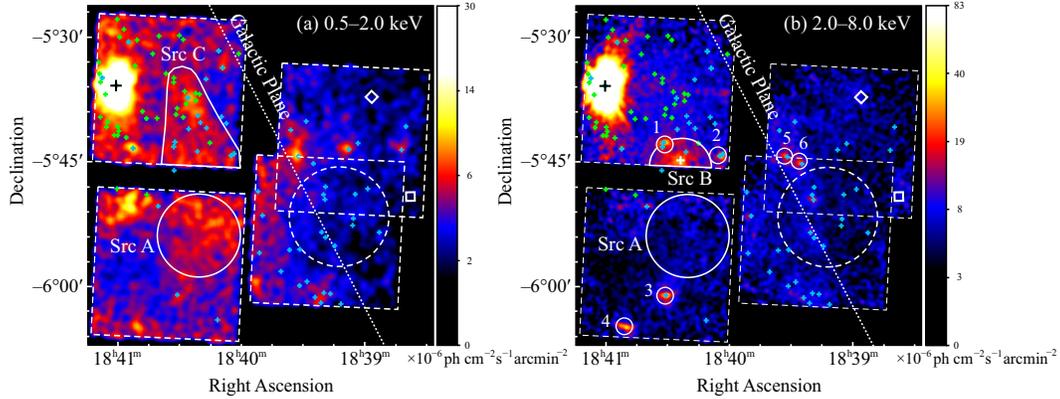}
\caption{(a) X-ray image in the energy band of 0.5--2.0~keV. Each dashed square shows the field of view of the XIS. The dotted line indicates the Galactic plane. A diffuse structure, 
Suzaku~J1840.2$-$0552
(Src~A), is marked with the solid circle with a $5'$ radius. 
The solid triangle-like region shows a complex emission (Src C) and positionally overlaps with G26.6$-$0.2 \citep{bamba03}. 
The brightest source marked with the black cross is 
the SFXT AX~J1841.0$-$0536 \citep{nobukawa12}. 
(b) Same as (a), but in 2.0--8.0~keV. The source indicated with the semi-ellipse  
with a major axis of $3'.6$ and a minor axis of $3'.1$
shows Suzaku~J1840.2$-$0544 or Src~B. It would be extended compared with the nearby point-like source numbered 1 (see text).
The white cross at the center of the Src~B region indicates the position of 3XMM~J184023.9--054445, which is an extended source (see text).
In the both images, the positions of point sources detected by {\it XMM-Newton} are indicated by small light green crosses (3XMM-DR4) and those detected by 
{\it Chandra} are indicated by small light blue crosses \citep{Ev10}. 
We extracted the spectra from the point sources numbered from 1 to 6. 
The background spectrum for these Suzaku sources was taken from the dashed circle. 
The white diamond symbol shows the position of the gamma-ray pulsar PSR~J1838--0537 \citep{Pl12} while
the white square symbol shows the position of the pulsar PSR~J1838-0549 \citep{hobbs04}.
}
\label{X-ray_image}
\end{center}
\end{figure}

In figure~\ref{X-ray_image}, we show  the X-ray images in the soft (0.5--2.0~keV) and hard (2.0--8.0~keV) bands. Vignetting of the XRTs and the difference of exposure times among the four observations are taken into account. 
The brightest source marked with the black cross in both of the images is the SFXT AX~J1841.0$-$0536 \citep{nobukawa12}. 
The light blue crosses indicate point sources which were detected by previous observations with {\it Chandra} \citep{Ev10}  
while the light green crosses are for {\it XMM-Newton}\footnote{http://xmmssc-www.star.le.ac.uk/Catalogue/}.

Although we find a largely-extended emission in the soft X-ray band, it is not clear only from figure~\ref{X-ray_image}a  
whether this extended emission is one simple diffuse source or a mix of diffuse emissions.
In fact, the extended emission in the northeast field may be contaminated by many point sources as are observed by {\it Chandra} and {\it XMM-Newton}. 
The southeast field is relatively free from point sources within the detection limit of {\it Chandra}. We, therefore, separately 
investigated a diffuse emission in the southeast field, Src A (section~\ref{srcA}) and that in the northeast field, Src C (section~\ref{NorthEast}).

\begin{figure}[htpb]
\begin{center}
\includegraphics[width=14cm]{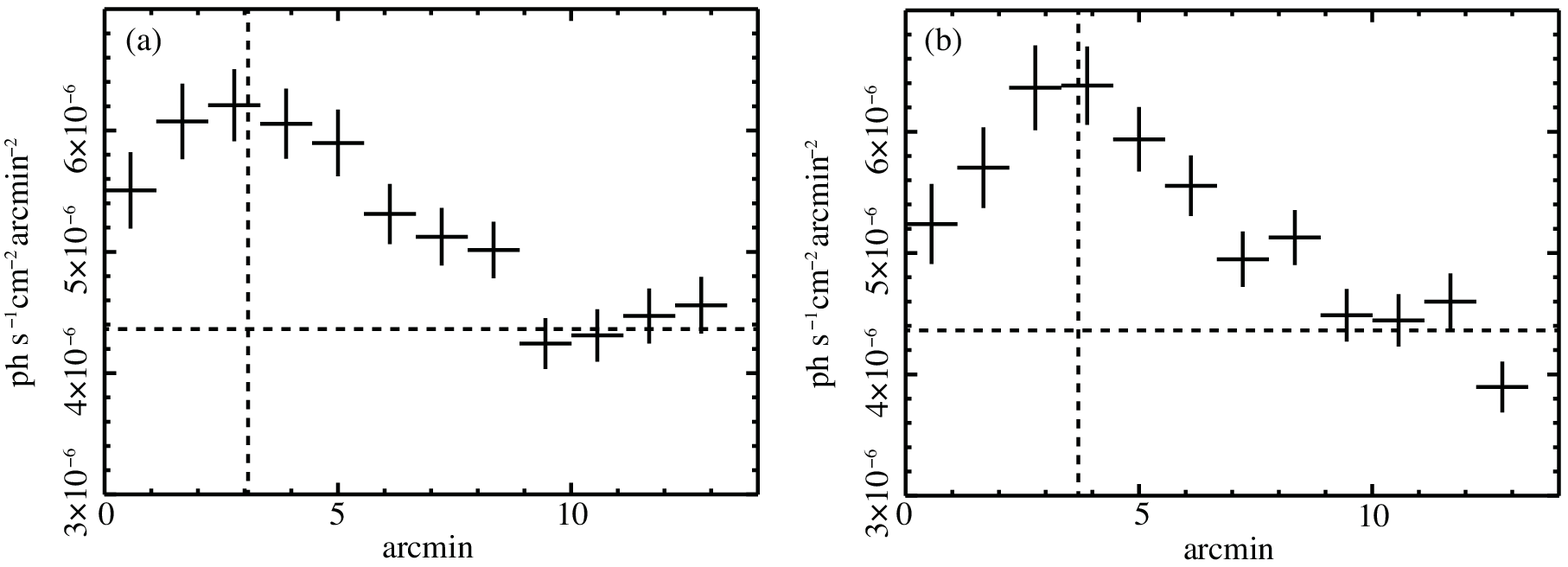}
\caption{
(a) Projection profile of Src~A along the right ascension after correcting the vignetting effect. 
The origin of the horizontal axis indicates the edge of the field of view. 
The vertical dashed line shows the center position of Src~A while the horizontal dashed line shows the constant level.
(b) Same as (a), but along the declination.
}
\label{radial_profile}
\end{center}
\end{figure}

To see the spatial structure of Src A, we show the projected profile in figure~\ref{radial_profile}.  We see a real extended structure by $\sim 10'\times10'$.  
The center position is $(\alpha, \delta)_{\rm J2000}=(18^{\rm h}40^{\rm m}11.^{\rm s}791, -5^{\circ}52'29.''83)$, and hence Src A is referred to as 
Suzaku~J1840.2$-$0552 (the solid circle in figure~\ref{X-ray_image}a). Whether Src A  is more extended to the northeast field (Src C) or not is unclear 
due to contamination of many point sources in Src C (see section~\ref{NorthEast}) 

No largely extended emission is found in the hard band (figure~\ref{X-ray_image}b). 
Instead, the source marked with the solid semi-ellipse in figure~\ref{X-ray_image}b (Src~B) seems to be slightly extended compared to the nearby point-like source 
numbered 1. The center position of the ellipse is $(\alpha, \delta)_{\rm J2000}=(18^{\rm h}40^{\rm m}24.^{\rm s}734, -5^{\circ}44'36.''83)$, 
and hence Src~B is referred to as Suzaku~J1840.2$-$0544.  This source is discussed in section~\ref{srcB}.

\subsection{The Galactic ridge background} 
\label{GRXE}

In the following spectral analysis for the largely diffuse sources, we used the background spectrum taken from the dashed-circle in figure~\ref{X-ray_image}, 
because no excess X-ray is found from this region. The background region is located on the Galactic plane at $(l,~b)\sim(26.^{\circ}3, 0.^{\circ}0)$ and is dominated by the Galactic ridge X-ray emission (GRXE; \citealt{Wo82, Ko86}).
\cite{Uchiyama2013} reported that the GRXE spectrum is represented by two optically-thin thermal plasma models with different temperatures ($kT=$1.3~keV and 6.6~keV) and a power-law ($\Gamma=2.13$) plus a neutral Fe~K$\alpha$ line at 6.4~keV from cold interstellar clouds or mediums in addition to the cosmic X-ray background (CXB) and the foreground thermal emission.

Using the GRXE model with the same parameters as those in \cite{Uchiyama2013} except for 
the absorption column density $N_{\rm H}$ and the normalization of each component, we successfully made the spectral fitting for the background ($\chi^2 /$d.o.f.$=212/201$). 
We obtained the absorption column density of $N_{\rm H}=(3.3\pm0.5)\times 10^{22}$~cm$^{-2}$.
This is consistent with those in other regions at $|l|\sim10^{\circ}$--20$^{\circ}$;
$N_{\rm H}= (3$--$4) \times 10^{22}$ cm$^{-2}$ at $(l, b)\sim(-10^{\circ}, 0^{\circ})\,$\citep{Yasumi14} and  
$N_{\rm H}= (2.2$--$2.5) \times 10^{22}$ cm$^{-2}$ at $(l,b)\sim(-23^{\circ}, -0.7^{\circ})$ \citep{Takata15}.

\subsection{Diffuse Soft X-ray source: Src~A} 
\label{srcA}

\begin{figure}[bpht]
\begin{center}
\includegraphics[width=8.5cm]{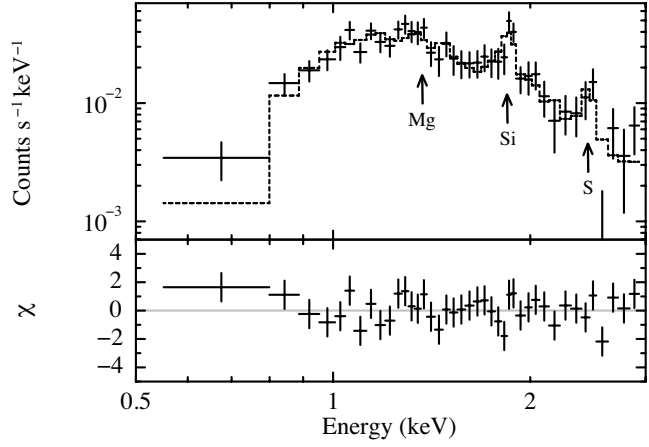}
\caption{(top) X-ray spectrum of Src~A. The Dashed line indicates the best-fit model described by an absorbed CIE plasma. (bottom) Residuals between the spectrum and the model.}
\label{src2_spec}
\end{center}
\end{figure}

The X-ray spectrum of Src A was made from the solid circle in figure~\ref{X-ray_image}a.
Figure~\ref{src2_spec} shows the background-subtracted spectrum. 
Since K$\alpha$ lines of Mg, Si, and  S are detected, the spectrum may be mainly dominated by an optically thin thermal plasma. 
We examined the spectrum with a model of absorbed collisional ionization equilibrium (CIE) plasma ({\tt vapec}$\times${\tt phabs} in XSPEC), and obtained an 
acceptable fit ($\chi^2$/d.o.f=35/35). 
The best-fit spectrum model is given by the dotted histogram in figure~\ref{src2_spec} while the best-fit parameters are summarized in table~\ref{src2fit}. 

\begin{table}[htbp]
\begin{center}
\caption{Best-fit parameters for Src~A.$^*$}
\begin{tabular}{llll}
\hline
Component 				& Parameter						&		& Value\\
\hline
\multicolumn{4}{l}{Model: {\tt vapec $\times$ phabs}} \\
\hline
Absorption$^{\dag}$ 	& Column density ($N_{\rm H}$) 	& 		& ($1.3\pm0.1)\times10^{22}$~cm$^{-2}$ \\
CIE plasma				& Temperature ($kT$) 			& 		& $0.7\pm0.1$~keV\\
			 			& Abundance (solar) 			& Mg	& $0.6^{+0.5}_{-0.4}$\\
						& 					 			& Si	& $1.0^{+0.5}_{-0.4}$\\
						& 					 			& S		& $1.1^{+0.9}_{-0.7}$\\
         				& Emission measure$^{\ddag}$    &       &  ($2.5\pm 0.2) \times 10^{-3}$~cm$^{-5}$ \\
\hline
\multicolumn{2}{l}{Flux (0.5--2.0~keV)}	    			& \multicolumn{2}{l}{$(3.7\pm0.2) \times 10^{-13}$~erg~s$^{-1}$~cm$^{-2}$} \\
$\chi^2/$d.o.f.			&								&		& 35/35\\
\hline
\label{src2fit}
\end{tabular}
\end{center}
$^*$Errors are at the 90\% confidence level. \\
$^{\dag}$Photoelectric absorption. \\
$^{\ddag}$Volume emission measure $\frac{10^{-14}}{4\pi D^2} n_{\rm e} n_{\rm H} V$, where
$D$, $n_{\rm e}$, $n_{\rm H}$, and $V$ are 
the source distance, electron and atomic hydrogen densities, and the plasma volume, respectively.
\end{table}

We extracted the X-ray spectrum from  the purely diffuse soft X-ray region (Src A) and ignored a possible extension to the northeast field (Src C),  
because the soft X-ray in Src C may be  contaminated  by many faint point sources of {\it Chandra} and {\it XMM-Newton}. Therefore, the obtained flux of Src A 
should be regarded as a lower-limit for a soft X-ray source. 

The flux of the Src A region (raw flux) in 0.5--2.0~keV is
$(8.1\pm0.2)\times10^{-13}$ erg~s$^{-1}$~cm$^{-2}$ while that of the background region is $(3.3\pm0.1)\times10^{-13}$ erg~s$^{-1}$~cm$^{-2}$. 
The diffuse excess of Src A is far larger than possible fluctuations of the GRXE. Moreover, the best-fit absorption column density ($N_{\rm H}$) is  
different between Src~A and the GRXE ($N_{\rm H} \sim 1.3\times 10^{22}$~cm$^{-2}$ and $3.3 \times 10^{22}$~cm$^{-2}$). 
Thus Src A is a real diffuse source with the significance level of 24~$\sigma$. 

On the other hand, we found no significant difference in the hard band flux (2.0--10~keV): the raw fluxes of the source and background regions are $(2.5\pm0.1)\times10^{-12}$ erg~s$^{-1}$~cm$^{-2}$ and $(2.4\pm0.1)\times10^{-12}$~erg~s$^{-1}$~cm$^{-2}$, respectively.

Since the absorption column density is less than that in the background spectrum of the GRXE,  Src~A may be located at the near-side of the Galactic ridge. 
Using the ratio of $N_{\rm H}$, we estimate the source distance to be $\sim$3--4~kpc. Then the electron density and thermal energy are 0.2 cm$^{-3}$ 
and $6\times10^{48}$~erg (as a lower-limit), respectively. 

\begin{figure}[htpb]
\begin{center}
\includegraphics[width=9cm]{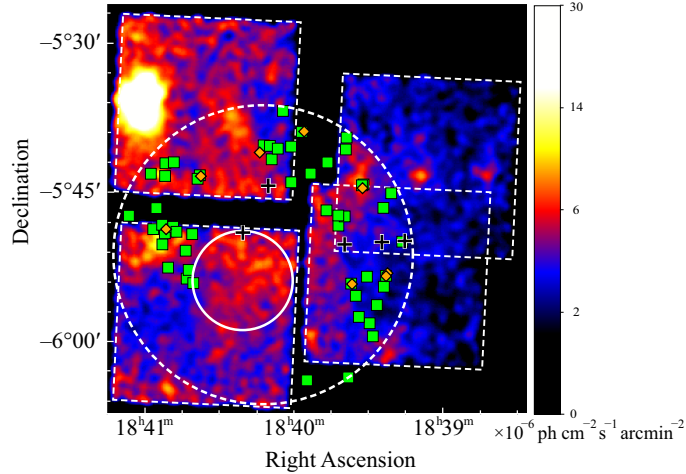}
\caption{Comparison between the X-ray image and  possibly related sources. The color scale is the same as figure~\ref{X-ray_image}a. The square, 
diamond, and cross marks show radio (continuum) sources, 
H\,{\small II} regions and B-type stars, respectively. The solid circle shows the Src A region while the dashed circle shows the $15'$ radius circle, where we searched the sources in the SIMBAD database.}
\label{src2_15arc}
\end{center}
\end{figure}

From the thermal spectrum, Src~A is likely either an SNR, an H\,{\small II} region, or a star cluster.
Using the SIMBAD database\footnote{http://simbad.u-strasbg.fr/simbad},
we found many radio sources and H\,{\small II} regions in the $15'$ radius circle around Src~A as shown in figure~\ref{src2_15arc}.  
However, no H\,{\small II} region coincides with the position and size of Src~A. No stellar cluster has been reported.

The X-ray plasma seems to be surrounded by the H\,{\small II} regions and the radio sources. 
Since five B-type stars exist in Src~A and its 15$'$ surrounding, it is presumed that several massive stars were produced in the past star formation at the cite. 
Src~A is therefore most likely an SNR that was produced by the explosion of a massive star. The thermal energy is also consistent with an SNR. 
Although no radio shell indicating the forward shock at the SNR surface has been reported so far, many radio sources may be confused with a shell structure, if any.

\subsection{Hard X-ray Source: Src~B} 
\label{srcB}
\begin{figure}[htpb]
\begin{center}
\includegraphics[width=8.5cm]{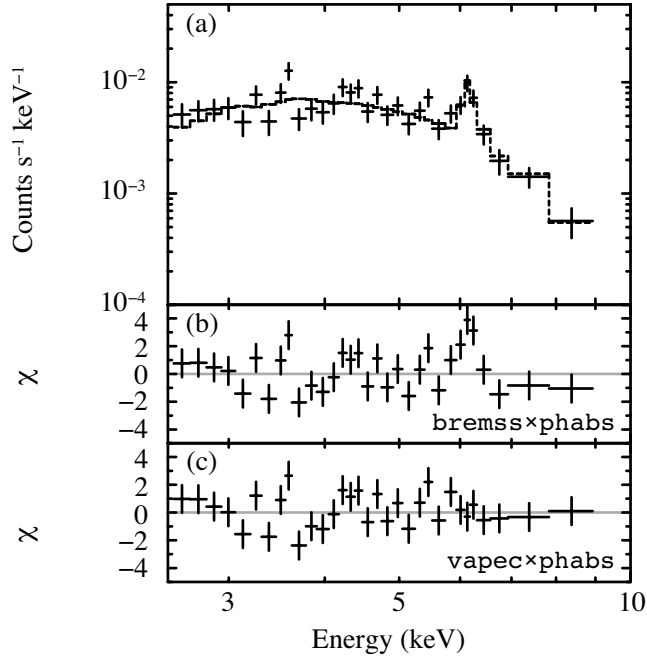}
\caption{(a) X-ray spectrum of Src~B. The dashed line is the best-fit model described by an absorbed CIE plasma ({\tt vapec}$\times${\tt phabs} in APEC) with 
the red-shift of $z=0.088$. (b) Residuals between the spectrum and the model consisting of {\tt bremss}$\times${\tt phabs}. (c) Same as (b), but the model consists of {\tt vapec}$\times${\tt phabs}.}
\label{src1_spec}
\end{center}
\end{figure}

The X-ray spectrum of Src B was made from a semi-ellipse with a major axis of $3'.6$ and a minor axis of $3'.1$. A nearby point source region 
(referred to as number 1 in figure~\ref{X-ray_image}b) was excluded.  
We made a background spectrum from the same region as Src~A (the dashed circle in figure~\ref{X-ray_image}b). 
Figure~\ref{src1_spec}a shows the background-subtracted spectrum for Src~B.

Neither a simple bremsstrahlung nor a power-law model with an interstellar absorption ({\tt brems}$\times${\tt phabs} or {\tt powerlaw}$\times ${\tt phabs}) gives 
a good fit with the reduced chi-squared values of 74/29 and 75/29, respectively.
The large residual is a line-like structure at $\sim 6.1$~keV (figure~\ref{src1_spec}b). 
We, therefore, added a narrow Gaussian line and obtained a better fit with $\chi ^{2}$/d.o.f.$={\bf 43/27}$.
The source and background fluxes in 2.0--10~keV are ($2.1\pm0.1) \times 10^{-12}$~erg~s$^{-1}$~cm$^{-2}$ and ($5.6\pm 0.1) \times 10^{-13}$~erg~s$^{-1}$~cm$^{-2}$, 
respectively. Thus the significance is 9~$\sigma$. The center energy and equivalent width of the Gaussian are $6.15\pm0.04$~keV and $430\pm130$~eV, respectively. 
The line detection is at the significance level of 5.7~$\sigma$. 

No atomic line at this energy in the rest-frame is found.
Since the most intense emission line in the 5--7 keV band is a K$\alpha$ line of Fe,
a possible origin would be a red-shifted Fe-K$\alpha$ line.
For a neutral Fe K$\alpha$ line at 6.4 keV, the redshift $z$ is 0.04 while for a He-like Fe K$\alpha$ line at 6.7~keV, $z$ is 0.09.

We finally fitted the spectrum with an absorbed CIE plasma ({\tt vapec}$\times${\tt phabs}) with a free parameter of a redshift (see figure~\ref{src1_spec}c). 
The redshift $z$ is found to be $0.088^{+0.008}_{-0.007}$. The best-fit parameters are summarized in table \ref{src1_fit}.

\begin{table}
\begin{center}
\caption{Best-fit parameters for Src~B.$^*$}
\begin{tabular}{lll}
\hline
Component & Parameter & Value\\
\hline
\multicolumn{3}{l}{Model: {\tt vapec $\times$ phabs}} \\
\hline
Absorption$^{\dag}$ 	& Column density ($N_{\rm H}$) 	& $6.7^{+2.1}_{-1.8}\times10^{22}$~cm$^{-2}$ \\
CIE plasma				& Temperature ($kT$) 			& $5.8^{+2.2}_{-1.4}$~keV\\
			 			& Fe Abundance (solar)			& $0.44^{+0.23}_{-0.15}$\\
						& Red-shift ($z$)				& $0.088^{+0.008}_{-0.007}$\\
\multicolumn{2}{l}{Flux (2.0--10~keV)}    				& $(1.5\pm0.1) \times 10^{-12}$~erg~s$^{-1}$~cm$^{-2}$ \\
$\chi^2/$d.o.f.			&								& 45/27\\
\hline
\end{tabular}
\label{src1_fit}
\end{center}
$^*$Errors are at the 90\% confidence level.\\
$^{\dag}$Photoelectric absorption. \\ 
\end{table}

The obtained absorption column density of $N_{\rm H}=6.7^{+2.1}_{-1.8}\times10^{22}$~cm$^{-2}$ is twice as large as that of 
the GRXE of $N_{\rm H}=$(3.3$\pm$0.5)$\times10^{22}$~cm$^{-2}$.
Thus together with the redshifted Fe-K$\alpha$ line,  
Src B is likely an extragalactic source; either a cluster of galaxies (CG) or an active galactic nucleus (AGN).
 
In order to distinguish these two possibilities, we investigate whether Src~B is extended or not. We made an intensity profile around Src~B along the right ascension 
in figure~\ref{X-ray_image}b, and fitted the profile with a Gaussian plus constant. The best-fit width of the Gaussian is $\sigma = 0'.91\pm 0'.10$ (at 1$\sigma$ error), 
which is significantly larger (by $3.1~\sigma$ confidence level) than the PSF size of $\sigma = 0'.58\pm 0'.04$ obtained from the nearby point source (numbered 1 in figure~\ref{X-ray_image}b).  Thus we conclude that Src~B is extended with the Gaussian size  of $0'.7\pm0'.1$ in $1\sigma$ error (after subtracting the PSF).

We also checked the {\it XMM-Newton} catalog and found an extended source as a counterpart of Src~B, 3XMM J184023.9$-$054445, whose position is indicated 
by a white cross mark at the center of Src B (figure \ref{X-ray_image}b). Its extension size is $0'.97 \pm 0'.05$ in the beta model, which supports our result.  
Thus Src B is likely a CG rather than an AGN. 

In addition, the observed equivalent width of  $\sim430$~eV is far larger than the typical value of AGNs (50--200 eV; \citealt{ricci14}) but is equal to that of CGs with the Fe abundance of 0.5~solar \citep{Fukazawa00, Matsushita11}.

\begin{table}
\begin{center}
\caption{Positions and absorption column density values of the three clusters of galaxies on the Galactic plane ($|l|<60^{\circ}$, $|b|<1^{\circ}$)}
\begin{tabular}{lll}
\hline
Source 		& Position                     & column density\\
          	&  ($l$, $b$)                  &  $N_{\rm H}$ ($10^{22}$~cm$^{-2}$) \\
\hline
XMMU~J183225.4$-$103645$^*$ 	& ($21^{\circ}.333$, $-0^{\circ}.658$)		& $7.9\pm0.5 $ \\
AX~J185905$+$0333$^\dag$	 	& ($36^{\circ}.974$, $-0^{\circ}.076$) 	& $9.0^{+1.2}_{-1.1} $\\
Suzaku~J1840.2$-$0544 (Src~B)	& ($26^{\circ}.566$, $-0^{\circ}.175$) 		& $6.7^{+2.1}_{-1.8}$ \\
\hline
\end{tabular}
\label{cluster_abs}
\end{center}
$^*$ \citet{nevalainen01}, $^\dag$ \citet{yamauchi11}.
\end{table}

A CG behind the Galactic plane with high dense gas cannot be observed in optical.
Only two CGs on the Galactic plane  ($|l|<60^{\circ}$, $|b|<1^{\circ}$) have been reported 
so far (\citealt{nevalainen01}; \citealt{yamauchi11}).  
Since a CG does not absorb X-rays by itself, the measured absorption column density provides that of the whole of our Galaxy.

We summarize the positions and absorption column density values of the three sources including this work in table~\ref{cluster_abs}. 
Although we have only three samples, we find that the closer to the Galactic center, the absorption column density increases.
Since more than ten thousand CGs have been discovered in the whole sky, 
more than one hundred CGs would exist behind the Galactic plane.
If the number of such samples increases, we will be able to measure the amount of matter through the whole of our Galaxy.

\subsection{Complex diffuse source Src C and Point Sources} 
\label{NorthEast}
In addition to Src A, the soft X-ray excess is found from the half region of the northeast field (Src~C: the triangle-like region in 
figure~\ref{X-ray_image}a). 
This position is almost overlapped with the hard X-ray source G26.6$-$0.1. 
\citet{bamba03} reported that the 0.7--10~keV flux of G26.6$-$0.1 is  $3.5\times10^{-12}$~erg~s$^{-1}$~cm$^{-2}$.

We made the X-ray spectra from Src~C and 6 point sources in the hard X-ray band (number 1--6 in figure~\ref{X-ray_image}b). The spectrum from Src~C includes 
the three discrete sources: Src~B, point source 1 and 2. The background spectrum is the same as that of Src~A and Src~B.
After subtracting the background spectrum, we fitted the spectra with an absorbed power-law and estimated the observed fluxes in the 0.5--2.0~keV and 0.5--10~keV bands. 
The results are listed in table~\ref{flux}. We also summarize in table~\ref{flux} the fluxes of Src~A and Src~B, and the integrated flux of point sources in the Src~C region 
detected by {\it Chandra} and {\it XMM-Newton}.

\begin{table}
\begin{center}
\caption{Summary of the observed flux for each source$^{*}$.}
\begin{tabular}{lcc}
\hline
Region 					& 0.5--2.0~keV			& 0.5--10~keV	\\
\hline
\multicolumn{3}{l}{{\bf Diffuse sources observed by {\it Suzaku}}} \\
\hline
Src~A				& $3.7\times10^{-13}$	& $5.6\times10^{-13}$\\
Src~B				& $7.2\times10^{-15}$	& $1.5\times10^{-12}$\\
Src C 				& $5.2\times10^{-13}$	& $3.6\times10^{-12}$\\
\hline 
\multicolumn{3}{l}{{\bf Integrated flux of point sources in Src C $^{\dag}$}} \\
\hline
& $1.1\times10^{-13}$	& $2.0\times10^{-13}$ \\
\hline
\multicolumn{3}{l}{{\bf Point sources detected by {\it Suzaku}$^{\ddag}$}} \\
\hline
point source 1		& $3.5\times10^{-14}$	& $6.3\times10^{-13}$	\\
point source 2		& $4.4\times10^{-14}$	& $5.6\times10^{-13}$	\\
point source 3		& $3.9\times10^{-14}$	& $4.8\times10^{-13}$	\\
point source 4		& $2.3\times10^{-14}$	& $5.3\times10^{-13}$	\\
point source 5		& $4.8\times10^{-14}$	& $2.5\times10^{-13}$	\\
point source 6		& $1.9\times10^{-14}$	& $3.9\times10^{-13}$	\\
\hline
\end{tabular}
\label{flux}
\end{center}
$^{*}$ The units are erg~s$^{-1}$~cm$^{-2}$. The interstellar absorption is not corrected.\\
$^{\dag}$ Point sources with {\it Chandra} and {\it XMM-Newton}. \\ 
$^{\ddag}$ See figure~\ref{X-ray_image}b for numbering.\\
\end{table}

In the soft X-ray band of 0.5--2.0 keV, the excess emission in Src C is $5.2\times10^{-13}$~erg~cm$^{-2}$~s$^{-1}$.
The summed flux of the resolved point sources in this region detected by {\it Chandra} and {\it XMM-Newton} is $1.1\times10^{-13}$~erg~cm$^{-2}$~s$^{-1}$. 
Thus no large contribution would be due to the integrated point sources, and a significant fraction of Src~C would be due to Src~A; 
Src~A may be elongated to the Src~C region as we noted in section~\ref{srcA}.

\subsection{Possible X-ray counterparts of HESS J1841$-$055} 
\label{HESS}
The X-ray flux (0.7--10 keV) of G26.6$-$0.1 observed by {\it ASCA} was reported to be $3.5\times10^{-12}$~erg~cm$^{-2}$~s$^{-1}$ \citep{bamba03}. This is 
consistent with the 0.5--10~keV flux of Src C. 
The flux of the integrated point sources observed by {\it Chandra} and {\it XMM-Newton} in the hard band is $2.0\times10^{-13}$~erg~cm$^{-2}$~s$^{-1}$.  
After subtraction of Src B and point source 1 and 2, the significant fraction of Src C (7.7$\times10^{-13}$~erg~cm$^{-2}$~s$^{-1}$) is due to diffuse 
emissions, which may be possible counterparts of HESS J1841$-$055. 
The flux ratio of gamma-ray and X-ray is $\sim40$ ($3.1\times10^{-11}$~erg~cm$^{-2}$~s$^{-1}$ for 
HESS~J1841$-$055; \citealt{aharonian08}). 
This ratio is consistent with a hadronic scenario \citep{yamazaki06}.

The brightest X-ray source in the HESS J1841$-$055 region is a SFXT, AX~J1841.0$-$0536. Since this source is a point source 
while HESS J1841$-$055 is largely extended, the SFXT would only locally and partially responsible for the high energy gamma-ray 
emission of HESS~J1841$-$055 \citep{sguera09}.  

Four pulsars are found near HESS~J1841$-$055. Two of them, PSR J1838$-$0549 and PSR J1838$-$0537, are in our X-ray images. 
PSR~J1838$-$0537 is a young gamma-ray pulsar \citep{Pl12} with the spin-down luminosity of $6\times10^{36}$ erg~s$^{-1}$. This pulsar, however, 
is largely offset from the main body of HESS~J1841$-$055, and hence could be at most partial (or local) origin of the TeV emission.
In addition, no significant X-ray is found from the positions of PSR J1838$-$0549 and PSR J1838$-$0537 \citep{Pl12} as shown in figure~\ref{X-ray_image}. 
The spin-down luminosity of PSR J1838$-$0549 is only $1\times 10^{35}$ erg~s$^{-1}$ \citep{hobbs04}, which is well below the required level for a counterpart of HESS~J1841$-$055.

\section*{Acknowledgement}
The authors thank all of the Suzaku team members for their full support of the Suzaku project. KKN is supported by Research Fellowships of Japan Society for the Promotion of Science for Young Scientists. This study was also supported by JSPS and MEXT KAKENHI Grant Numbers 24740123 (MN), 20340043, 23340047, 25109004 (TGT), and 24540229 (KK).



\end{document}